\documentstyle[12pt,aasms4]{article}

\begin{document}

\title{Dynamical Creation of Channels for Particle Escape \\
in the Solar Corona}
\author{Swadesh M. Mahajan\altaffilmark{1}}
\affil{Institute for Fusion Studies, The University of Texas at
Austin, Austin,Texas 78712}
\author{Riszard Miklaszewski\altaffilmark{2}}
\affil{Institute of Plasma Physics and Laser Microfusion, 00-908
Warsaw, Str. Henry 23, P.O. Box 49, Poland}
\author{Komunela I. Nikol'skaya\altaffilmark{3}}
\affil{Institute of Terrestrial Magnetism, Ionosphere and Radio Wave
Propagation, Troitsk of Moscow Region, 142092, Russia}
\author{and \\
Nana L. Shatashvili\altaffilmark{\,4,0} }
\affil{High Temperature Plasma Center, The University of Tokyo, Tokyo
113-0033, Japan}
\altaffiltext{0}{\small {\it Permanent address:} \ Plasma Physics
Department, Tbilisi State University, Tbilisi 380028, Georgia}
\altaffiltext{1}{\small Electronic mail: \ mahajan@mail.utexas.edu}
\altaffiltext{2}{\small Electronic mail: \ rysiek@ifpilm.waw.pl}
\altaffiltext{3}{\small Electronic mail: \ knikol@sdpl.izmiran.rssi.ru}
\altaffiltext{4}{\small Electronic mail: \
nana@plasma.q.t.u-tokyo.ac.jp
\hskip 0.3cm nanas@iberiapac.ge}

\clearpage

\begin{abstract}
It is shown that the connection of sufficiently fast flows with
dynamical channels for particle escape in the Solar Corona is
rather direct: it depends on their ability to deform (in specific
cases to distort) the ambient magnetic field lines to temporarily
stretch (shrink, destroy) the closed field lines so that the flow
can escape the local region. Using a dissipative two--fluid code
in which the flows are treated at par with the currents, we have
demonstrated channel creation in a variety of closed--field line
structures. This self--induced transparency constitutes the active
mode for the formation of the solar wind.
\end{abstract}


\clearpage

\clearpage

\section{Introduction}

\label{sec:intro}

The knowledge of the structure of the coronal magnetic field (see e.g.
\cite{lin}) serves as the principal guide in the construction of all modern
theories on the origin of the Solar Wind (SW), a stream of high speed
charged particles that manages to escape the solar atmosphere. There are
two possible modes of escape from the magnetic field: 1) the passive mode in
which the particles find a region of open field lines and escape without
affecting the ambient field, and 2) the active mode in which the particle
flows are strong enough to distort/modify the local magnetic field to create
their own escape channels. The former would have been a sufficient solution
if the wind emergence was limited to the heliographic polar regions,
especially at solar minima when the open field line structures are
relatively stable lasting several years. The observational story, however,
is very different. Habbal and Woo (2001), for example, have shown (by a
careful and systematic comparison of the ULLYSES data with coronal
measurements at $1.15\,R_{\odot}$) that the fast solar wind detected by
ULLYSES seems to arrive (mostly radially) from all latitudes of the
so--called quiet Sun. The passive mode scenario, then, would imply regions
of open magnetic flux over a large fraction of the solar surface --- an
implication that runs counter to observations.

From the most detailed coronal images available to date e.g., from
TRACE (\cite{schrijver}), the diffuse quiet Sun seems to be
studded with a gamut of loop--like structures; no open magnetic
field lines can be distinguished in these images. Even in the so
called coronal holes (low temperature regions originally believed
to have open field lines) there is no direct evidence for open
magnetic field lines. For example, (\cite{chertok}) recently
reported the disappearance/eruption of a "reasonably large
filament  located within an extended trans--equatorial CH.  The
disappearance was accompanied by a number of large--scale dynamic
phenomena such as a coronal mass ejection (CME), EUV--emitting
structures inside the CH, a soft X--ray arcade, and other effects.
These features seem to mean that CH structures need not be so
simple and quiet as previously assumed ... and can contain local
areas of closed large--scale magnetic fields, at least at low
altitudes". The authors also claim that "consideration of the {\it
Yohkoh}/SXT and {\it SOHO}/EIT data for other periods shows that
similar but not so spectacular CH--interior filaments and extended
accompanying activity can be found in some other events".
Indication of complicated magnetic topology and fine--scale
structuring of corona including CH--s can be found in
two--temperature coronal models from  {\it SOHO}/EIT observations
(see e.g. \cite{zhang}).

Information about the magnetic field is hard to extract at
chromospheric heights because of the low sensitivity and lack of high
spatial resolution of the measurements coupled with the inhomogeneity
and co--existence of small-- and large--scale structures with
different temperatures in nearby regions (\cite{A1}). In \cite{zhang}
it was shown that: (i) while the raw EIT images are dominated by the
spatial distribution of emission measure in the corona, the
temperature maps often emphasize fine structure, which is less
visible in the flux images; (ii) the emission measure of the hot
component is always found to be at least as large as that of the cool
component, meaning that single--temperature models miss most of the
coronal plasma to which EIT is sensitive. Also, it was shown that
blend of very cool Mg/Si lines with hot Fe xv lines in the EIT 284
$A^{o}$ bandpass induces a false solution for the hot component in
CH-s. Thus, such models were found unsatisfactory in CH-s.

Observations seem to suggest, then, that the genesis of the solar wind may
lie in the active mode of particle escape. If a given stream of particles
were to punch out its own channels of escape in a short--lived, dynamic
process, we could certainly explain the emergence of the wind from regions
of the solar surface with no observable long--lived (quasi--static)
open--field line structures; the flow enters a closed field line
region (preferably with weak fields), quickly distorts it, creates a
channel, escapes and leaves the field lines to mend themselves. This
kind of phenomenon will happen with statistical uniformity over the
entire solar surface and the wind would appear to come from the
regions permeated by primarily closed field line structures.

In this paper we report preliminary results demonstrating channel creation
by sufficiently strong flows in regions of relatively weaker fields (\cite
{woo}). We are motivated by the mounting evidence that strong flows are
found everywhere --- in the sub--coronal (chromosphere) as well as in the
coronal regions (see e.g. \cite{schrijver,golub,wilhelm,A1,A2,ami1,ami2,debi}
and references therein), and by the growing belief and realization that the
plasma flows may complement the abilities of the magnetic field in
the creation of the amazing richness observed in the coronal
phenomena (\cite{MMNS2}).  In the latter paper, a short survey of the
published data on flow evidence was given. We would repeat here the
most representative specimens.

There is an abundance of short-lived (minutes) jets observed at the
base of corona (\cite{b1,b2,w1,w2,woo2}) ; these jets can  play a
crucial role in the creation of Coronal structures (including CH-s).
``When observed with high spatial resolution, the atmosphere at the
base of the corona is found to be dynamic with a large fraction of
the surface covered by chromospheric and transition region material
moving up and down in structures with characteristic sized of
$\sim 1000\,km$. Three types of jets (cold, with temperature $\sim 10^4\,K$)
are observed: the ubiquitous spicules; macrospicules, large spicules best
observed in CH-s and
high speed jest with velocities of the order of $400\,km/s$". It
should be mentioned that at present the knowledge of spicules is
rather poor. ``The upward mass flux provided by spicules is
approximately 100 times that lost in the solar wind outflow. Hence,
most of this mass, if heated to coronal temperatures, must cool and
fall back into the chromosphere, with only a small fraction being
carried out in SW. However, it is not known what fraction of the mass
observed in spicules gets heated to coronal temperatures. There is no
direct evidence that significant amounts of spicular material are
heated to coronal temperatures, and some limited evidence that most
of the mass in spicules remains at lower temperatures" (\cite{w2}).
It is estimated that the energy flux carried by macrospicules is
$\sim 2\cdot 10^7\,erg/cm^2s$ while the smaller spicules account
for fluxes that are an order of magnitude smaller. Assuming that
they appear spontaneously in time and space, and knowing from the
latest observation that the corona is very diverse
and dynamical (with continual appearance/disappearance of bright
structures in specific regions, though the corona as a whole shines
forever) one could certainly count the up--flows in the macrospicules
and large spicules as possible energy and material sources for the
coronal structures and the solar wind. At the end we quote from
(\cite{schrijver}) on the TRACE observations: "the EUV observations
by TRACE reveal not only material at coronal temperatures moving
upward from as low as a few thousand kilometers above the
photosphere, but also cool material, no hotter than about
$20000\,K$".

Detailed exploration of the mechanisms which generate the observed
flows is beyond the scope of this paper. But, guided by
phenomenology, we do give here a few examples: based on the
estimates of energy fluxes required to heat the chromosphere and
Corona, Goodman (2001) has shown that the mechanism that
transports mechanical energy from the convection zone to the
chromosphere (to sustain its heating rate) could also supply the
energy to heat the corona and accelerate the Solar wind. The
heating/acceleration problem is, thus, shifted to the problem of
dynamic energization of the chromosphere for which the flows are
found to be critical as warranted by the observations made by
TRACE: the overdensity of coronal loops, the chromospheric upflows
of heated plasma, and the localization of heating function in the
lower corona (\cite{A1,A2,schrijver} and references therein).
Catastrophic models of flow production in which the magnetic
energy is suddenly converted into bulk kinetic energy (and thermal
energy) are rather well--known; various forms of magnetic
reconnection (flares, micro and nanoflares) schemes permeate the
literature (see e.g.~ Wilhelm 2001; Christopoilou, Georgakilas and
Koutchmy 2001 for chromosphere up--flow generations). A few other
mechanism of this genre also exist: Uchida et. Al. (2001) proposed
that the major part of the supply of energy and mass to the active
regions of the corona may come from a dynamical leakage of
magnetic twists produced in the sub-photospheric convection layer;
Ohsaki, Shatashvili, Yoshida and Mahajan (2001,2002) have shown
how a slowly evolving closed structure (modeled as a Double
Beltrami two fluid equilibrium) may experience, under appropriate
conditions, a sudden loss of equilibrium with the initial magnetic
energy appearing as the mass flow energy. Another mechanism, based
on loop interactions and fragmentations and explaining the
formation of loop threads, was given in Sakai and Furusawa (2002).
More steady mechanism of chromospheric flow generation was given
in \cite{MNSY}. The dynamic 2D modeling of the same mechanism
constitute the subject of a future submission.

The magneto--fluid coupling that will be shown to lead to channel--creation
has been the subject of intense theoretical study in the last few years. A
simple two--fluid model with arbitrary flows has been explored to reveal
the breadth of phenomena made possible by the combined actions of the
flow--velocity and the magnetic fields. A remarkable evidence of the
magnetofluid coupling ``in action'' is the recent demonstration that
the hot coronal structures could be created from the evolution and
re--organization of a relatively cold plasma flow (coming from the sub--coronal
region) in the presence of the ambient magnetic field anchored inside the solar
surface (\cite{MMNS1,MMNS2}). The heating of the structures takes place due
to the viscous dissipation of a part of the flow kinetic energy during the process
of particle trapping and accumulation; this happens in regions of relatively
strong magnetic fields.

In this paper we wish to demonstrate that this very interaction between the
flows and the magnetic fields provides the crucial ingredient in the physics
of channel creation. What we need in this case are relatively stronger flows
pushing their way through regions of relatively weaker fields. The detailed
nature of a channel will depend on the initial, and the boundary conditions.
Simultaneously with the creation of escape--channels, plasma heating takes
place both in the created channels and in their neighborhood. Viscous dissipation,
the primary heating mechanism in the model, preferentially heats ions --- a
feature essential to reproduce what is observed in the open field--line
regions (see e.g. \cite{bravo,marsch1,marsch2}). Although primary heating
is an integral part of the model, we will not dwell on it and limit this
paper to elucidating the fundamentals of channel--creation. In fact, for
simplicity, we will choose equal electron and ion temperatures.

\section{Model}

\label{sec:Model}

The physical model investigated for channel creation is a simplified
two--fluid model. The results presented here are obtained from a 2
dimensional (2D) simulation code reported in (\cite{MMNS2}). We use
quasi--neutrality --- electron and proton number densities are nearly
equal: $n_e\simeq n_i = n$ ($\nabla \cdot {\bf j} = 0$) but allow the
electron and the proton flow velocities to be different. Neglecting
electron inertia, these are $V_i$ and $V_e=(V-j/en)$, respectively.
We assign equal temperatures to the electron and the protons so that
the kinetic pressure $p$ is given by: $p = p_i + p_e \simeq 2\,nT , \
T = Ti\simeq Te $. The analysis can be readily extended later to the
more realistic case of different temperatures for different species (see
e.g. \cite{swT}). We are aware of several studies of the solar wind
problem using multi--fluid, multi--dimensional descriptions (see e.g. \cite
{marsch1,marsch2,hollweg} and references therein) that even include the
self--consistent effects of MHD waves on minority ions. We however,
believe, that essential features of the primary flow--based physics
of escape channels can be captured with our basic model.

The dimensionless two-fluid equations describing the flow-field
interaction processes can be read from \cite{MMNS1}:
\begin{equation}
\frac{\partial }{\partial t}{\bf V}+({\bf V}\cdot \nabla ){\bf V}=\frac{1}{n}%
\nabla \times {\bf b}\times {\bf b}-\beta \frac{1}{n}\nabla (nT)+\nabla
\left( \frac{r_{A}}{r}\right) +\nu _{i}(n,T)\left( \nabla ^{2}{\bf V}+\frac{1%
}{3}\nabla (\nabla \cdot {\bf V})\right) ,  \label{eq:motion}
\end{equation}
\begin{equation}
\frac{\partial }{\partial t}{\bf b}-\nabla \times \left( {\bf V}-\frac{%
\alpha }{n}\nabla \times {\bf b}\right) \times {\bf b}=\alpha \beta \ \nabla
\left( \frac{1}{n}\right) \times \nabla (nT),  \label{eq:field1}
\end{equation}
\begin{equation}
\nabla \cdot {\bf b}=0,  \label{eq:field2}
\end{equation}
\begin{equation}
\frac{\partial }{\partial t}n+\nabla \cdot n{\bf V}=0,
\label{eq:cont}
\end{equation}
\begin{eqnarray}
\frac{3}{2}n\frac{d}{dt}(2T)+\nabla ({\bf q}_{i}+{\bf q}_{e}) &=&-2nT\nabla
\cdot {\bf V}+2\beta ^{-1}\nu _{i}(n,T)\,n\left[ \frac{1}{2}\left( \frac{%
\partial V_{k}}{\partial x_{l}}+\frac{\partial V_{l}}{\partial x_{k}}\right)
^{2}-\frac{2}{3}(\nabla \cdot {\bf V})^{2}\right]  \nonumber \\
&&+\frac{5}{2}\alpha (\nabla \times {\bf b})\cdot \nabla T-\frac{\alpha }{n}%
(\nabla \times {\bf b})\nabla (nT)+E_{H}-E_{R}.  \label{eq:heat}
\end{eqnarray}
where the notation is standard with the following normalizations: the
density $n$ to $n_{0}$ at some appropriate distance from the solar
surface, the magnetic field to the ambient field strength at the same
distance, and velocities to the Alfv\'{e}n velocity $V_{A0}$. The
parameters $r_{A0}=GM_{\odot }/V_{A0}^{2}R_{\odot }=2\beta
_{0}/r_{c0},\ \alpha _{0}=\lambda _{i0}/R_{\odot },\ \beta
_{0}=c_{s0}^{2}/V_{A0}^{2}$ are defined with $n_{0},\ T_{0},\ B_{0}$.
Here $c_{s0}$ is a sound speed, $R_{\odot }$ is the solar radius,
$r_{c}=GM_{\odot }/2c_{s0}^{2}R_{\odot }$, $\lambda _{i0}=c/\omega
_{i0}$ is the collisonless skin depth, $\nu _{i}(n,T)$ is ion kinematic
viscosity and $q_{e}$ and $q_{i}$ are electron and ion dimensionless heat flux
densities, $E_{H}$ is the local mechanical heating function and
$E_{R}$ is the total radiative loss. We note that the full viscosity
tensor relevant to a magnetized plasma is rather cumbersome, and we
do not display it here. Just to have a feel for the importance of
spatial variation in viscous dissipation, we display its relatively
simple symmetric form. It is to be clearly understood that this
version is meant only for theoretical elucidation and not for
detailed simulation. We notice that even for incompressible and
currentless flows, heat can be generated from the viscous dissipation
of the flow vorticity. For such a simple system, the rate of kinetic
energy dissipation turns out to be
\begin{equation}
\left[ \frac{d}{dt}\left( \frac{m_{i}{\bf V}^{2}}{2}\right) \right] _{{\rm
visc}}=-m_{i}n\nu _{i}\left( \frac{1}{2}(\nabla \times {\bf V})^{2}+\frac{2}{%
3}(\nabla \cdot {\bf V})^{2}\right) .  \label{eq:visc}
\end{equation}

What are the conditions for the dissipation rate to be large enough
for  effective plasma heating? This and the related question of the
requirements on the  radial energy fluxes for the flow--based
mechanisms to be meaningful (for coronal heating and  generation of
the solar wind) were  examined in \cite{MMNS2}. The flow requirements
were found to be quite  consistent with the latest observational data.
It was, however, shown that in the absence of ``anomalous viscosity",
the only way to enhance the dissipation rates (to the observed values)
through viscosity is to create spatial gradients of the velocity field
that are on a scale much shorter than that of the structure length
(defined by the smooth part of the
magnetic field). Thus, the viability of this two--fluid  approach
depends wholly on the existence of mechanisms that induce short--scale
velocity fields. Theoretical foundations taking into account the fundamental
role of Hall term and numerical simulations (without Hall term) showed that
the short--scale velocity fields are, indeed, self--consistently generated
in the two--fluid system.

\section{Deformation of closed field lines}

A high--speed flow in or near the transition region (TR) must overcome
both gravity and the magnetic field to emerge as the solar wind. Overcoming
gravity, by itself, imposes a stiff lower bound on the flow velocity.
Negotiating the magnetic field is even harder; preliminary studies show
that flows with reasonable TR densities and velocities $\leq 400\,km/sec$ can
not destroy or deform closed magnetic fields structures sufficiently
to meet escaping conditions. Estimates based on the observed magnetic field
strengths show that even in weak field regions ($\sim (1-5)\,G$) flows must
be rather strong to punch holes in the structure. If the up--flow creation
and acceleration mechanisms were operative somewhere below the hot corona,
the flow--magnetic field interaction could lead to conditions more favorable
to particle--escape.

We want to remind the reader that there is an implicit assumption in
this paper that the short-lived processes like flares, CME in inner
corona (explained by catastrophe models) are characteristic for the
active Sun and can not give continuous material and energy supply to
relatively permanent particle escape process from all over the solar
surface during its entire cycle (including quiescent period).

As mentioned earlier the creation of flows is a major subject beyond
the scope of this paper. For some of the recent theoretical work on
chromospheric--TR  acceleration/flow--generation, the reader may consult:
(\cite{marsch1,marsch2,hollweg,ofman,kohl,elements}) for most promising SW
acceleration and heating models based on high--frequency Alfv\'en waves,
and (\cite{osym2, MNSY}) based on magneto--fluid coupling.

Some of the recent observational  findings on the solar wind are also
highly revealing.  \cite{dflsw}, analyzing SUMER measurements
(SOHO) on the polar CHs, found that the non--thermal motions
sometimes, but not always, increase slightly with height above the
limb. Based on the Doppler shift measurements of \cite{mariska} using
SUMER coronal hole and quiet Sun Spectra, they speculate that this
may be a manifestation of the fast Solar wind. They also report that
"Cooler plasma may be trapped in closed structures and not
participate in the flow, or the flow may not begin in open structures
until the temperature reaches values near $6\cdot 10^5\,K$". At the
same time IPS observation (EISCAT and VLBA systems of radio wave
telescopes) carried out specifically for measuring the fast SW speed
as near to the Sun as possible deduced average speeds $\sim
800\,km/s$ with very large scattering within heliocentric distances
$2 - 10\,r_{\odot}$ (\cite{od,nature})

In the light of the preceding discussion, we shall simply assume that high
speed flows are already there below the coronal base where they begin to
nteract with the closed field regions; they provide the initial conditions in
our numerical work. We shall also, justifiably, assume that the processes
that generate the primary flows and the primary solar magnetic fields are
independent (say at $t=0$). Simulation of two distinct representative problems
will be presented: 1) the flow interacting with a single structure providing
the simplest example of field--deformation, and 2) the flow passing through
a multiple structure region creating escape--channels under specific conditions.

The numerical code to solve (1-5) was constructed in 2D flat geometry (x,z)
using the 2D version of Lax--Wendroff numerical scheme (Richtmyer and
Morton 1967) alongwith applying the Flux--Corrected--Transport procedure (\cite{rm}).
Equation (2) was replaced with its equivalent for the y--component of the vector
potential which automatically ensures the divergence-free
property of the magnetic field. The equation of heat conduction was treated
separately by Alternate Direction Implicit method with iterations
(\cite{zalesak}). Transport coefficients for heat conduction and
viscosity were taken from Braginski, 1965.  We were quite careful in
choosing the radiation loss term ($E_{Br}$ denotes Bremsstrahlung radiation),
\ $E_R=2\cdot E_{Br}=2\cdot 1.69\cdot 10^{-25}\cdot n^2\cdot T^2Z^3
erg/cm^3s$, with $\  Z=1$ \ ; The choice was  based on the results of
\cite{rtv,ct,pot,tk,mtw} (\cite{MMNS2}). And since we have a built--in heating
mechanism no external heating source $E_H$ was invoked though we believe that
in the escaping channels one can later add secondary events like wave
generation for  additional acceleration mechanisms.  A numerical
mesh of \ $200\times 150$\  points was used for computation.

The initial solar magnetic field, following \cite{MMNS2}, was modelled as a
2D arcade with circular field lines in the $x$--$z$~plane (see first plot of
Fig.~1b for the contours of the vector potential, or the flux function). The
field attains its maximum value  $B_{\rm max}(x_o, z=0)$ at $x_0$ at the center
of the arcade, and is a decreasing function of the height $z$ (radial
direction).

Note that the 2D Cartesian nature of our code does not allow us to explore
large distances from the surface due to interference with the boundaries.
Although we present here only the symmetric cases, the simulation of
asymmetric situations is straightforward.

\subsection{Deformation of single closed field structure}

We first study the dynamics of a spatially localized flow (taken to be
initially a Gaussian as motivated by observations - e.g. jets, spicules)
entering an arcade--like single closed field line structure. Two palpably
distinct scenerios emerge:

1) When the flow is strong ($|{\bf V}_0|_{max}\sim 600\,km/sec, \ n_0\sim
10^8\,cm^{-3}$) and its peak is located in the central region of the
arcade magnetic field structure, the original field ($B_{0max}=5\,G$)
shown in (Fig.1a) is seriously deformed (see Fig.1b   representing
the evolution of the arcade-like magnetic structure deformation
process for three time--frames: $t=0; \ 768, \ 1749\,sec$), and its
central region is transformed to one with more or less parallel field
lines. The local channel, however, does not go all the way but may
extend to a respectable height. The resulting plasmoid--type
configuration, though,   may not lead to the particle escape. In all
such cases one finds that narrower the flow pulse, the sharper the
shear created, and stronger the flow, the faster is the
deformation process.

2) When several flow pulses arrive simultaneously towards a single
arcade structure, they may create, in the central region, sheared narrow
sub-regions with opposite polarity . The magnetic ``well'' displayed in
Fig.2 (Fig.2b, again, displays the deformation process for three
time--frames: $t=0; \ 768, \ 1749\,sec$), for instance, was formed by
two identical pulses (see Fig.2a, \ $|{\bf V_0}|_{max}\sim
600\,km/sec, \ n_0\sim 10^8\,cm^{-3} $) located symmetrically on the
opposite sides of the arcade-center with ($B_{0max}=5\,G$).

In carrying out the simulations an important assumption was made, namely,
the diffusion time of magnetic field is longer than the duration of the
interaction process (it would require the plasma temperature to be at least
a few $eV$--s.)

In both of the above cases, the flows were not able to punch escape--channels
although the ambient field was quite thoroughly deformed. This is true even
when we put somewhat larger but realistic amount of energy in the flows;
the flow cannot overcome the magnetic field in a direct ``collision'.

Note that we are not discussing here a possible pathway for the creation of
cold prominences; we believe that a 3D dynamical picture is essential for
the study of such events. In addition, this paper deals only with
relatively strong flows since the weaker flows are expected to be
trapped in the structure and will, perhaps, create hot and bright
areas near the TR--coronal interface without any serious deformation
of field--lines \cite{MMNS2}.

\subsection{Deformation of neighboring closed field structures --
Channels for particle escape}

We have just seen that the direct attempt by a relatively strong flow to
force its way through a moderate--strength single magnetic field structure
resulted in complete failure; the field is highly distorted but does not
quite yield. We must, therefore, look for other magnetic field configurations
(prevalent in the corona) which might be more cooperative. Recent literature
is extremely helpful in this quest. It has been suggested in \cite{woo} that
the coexistence of strong-- and weak--field components observed in the quiet--Sun
photospheric field (Lin (1995) and Lin \& Rimmale (1999)), and supported by
theoretical investigations of the solar dynamo (e.g., Cattaneo 1999), has a
counterpart in the corona. Through this study it was shown that the
observed predominance of the radial component of the quiet coronal
magnetic field is defined, again, by the weak--field component.

Coupling these observations with the models of high--speed up--flow
generation in the chromosphere and corona, it seems rather reasonable
to study the passage of a strong flow through multi--arcade magnetic
field structures. Although it is only the 3D simulations that can
reproduce most of the observational features of the channel escape
process, we believe that the current 2D code is sufficient to prove
or disprove the principal point, that is, whether an escape--channel
can be created.

We describe two representative case studies: 1) the flows interacting with
two neighboring arcade--structures, and 2) the flows interacting with four
neighboring arcades. For optimum effect we locate the maxima of the flow
pulses in the weak--field region in between the neighboring arcades. For
this study, the arcade structure is taken to be symmetric. It must be
stressed that inhomogeneous initial conditions do lead to different
evolutions, but channel creation remains a common feature.

The fate of the two--arcade structure when invaded in the middle (weak
Field regions ($B_{0max}=3\,G$)) by a fast flow ($|{\bf V_0}|_{max}\sim
900\,km/sec, \ n_0=2\cdot 10^7\,cm^{-3}$) is shown in Fig.3. The flow is
able to stretch and drastically deform the structure and, in a reasonably
short time ($\sim 50\,min$), create a channel for escape (see the last
time--frame of Fig.3b). The channel itself is practically cold for
distances of a few $R_{\odot}$. The neighboring regions are
comparably hotter: at the coronal base (created in this dynamical
process), one can distinguish rather hot ($T\sim 10^6\,K$) areas
where a part of the flow was trapped and thermalized  (the heating
source and process, described in \cite{MMNS2} and  discussed in the
introduction, is due to the effective viscous dissipation of the
short scale shocks created in the velocity field). Note, that if the
simulation were done in cylindrical $(r,\phi )$ geometry, we could
see the widening of the channel with increasing $r$.  We would also
add here that if the short scales created in the velocity fields due
to Hall term and vorticity effects, and secondary processes like wave generation in the channel
were also taken into account one could get  hotter channels of escape.
We are refining as well as modifying our model to incorporate these processes.

One of the more interesting consequences of the channel--creation dynamics
is displayed in (Fig.4) ---there is a sharp decrease in density (Fig.4(a))
along the channel (after the usual shock front area due to the interaction
of flow with background plasma) with a clearly distinguishable ballistic
deceleration of the flow (Fig.4(b)). At heights $\geq2\,R_{\odot}$
from the Sun's surface the flow speed ($\sim 800\,km/sec$) has only
marginally decreased ; the fast flow expends a negligible fraction of
its energy in creating a channel for its escape; the escaping flow is
almost as fast as the entering flow.

The response of a 4-arcade structure to the onslaught of the flow is rather
inhomogeneous and complicated; several channels are created in the region
of the flow (Fig.5). The central channel seems a bit pressed due to combined
interactions (Fig.5(c))- but this could be just an artifact of the Cartesian
geometry used here. In the dynamical evolution of this system, there is
a ballistic deceleration of the flow in each one of the channels; the
deceleration is faster in the central channel (see Fig.5(c)). One can also
see that, at longer times, the three structures will be permeated by flows.
This picture could be seen as a possible depiction of the complex and very diverse dynamical
structure of the recently observed  Coronal Holes.

One word of caution is necessary : for all our runs, we assumed
that the flows were, initially, constant in time. We understand
that up--flows from the chromosphere or TR have finite life times;
the channel creation process, therefore, will last only for the
time dictated by the duration and other characteristics of the
impinging flow--pulse.

We now list several other omissions of this preliminary study:
Anisotropies of velocities and temperature (source of wave generation
and instabilities), ionization, multi--species dynamics, flux
emergence etc. are not included. Either of these could influence the
channel--creation dynamics. We, however, believe that our simple
model has adequately shown that sufficiently strong flows are capable
of engineering their escape (self--induced transparency)
from a variety of closed field line structures prevalent in the solar
atmosphere.

\section{Conclusions}

\label{sec:conclusions}

By suggesting and investigating an active pathway by which strong flows can
create temporary channels (with practically radial open field lines)
through which they can escape the combination of gravity and ambient magnetic
fields, this paper advances a possible resolution of the observational dilemma
--- that the Solar wind seems to originate (more or less uniformly)
from the entire solar surface believed to be studded with closed field line
structures. The intense re--arrangement of the magnetic field lines needed
for the formation of escape channels is a consequence of the magneto--fluid
coupling --- the ability of a strong flow to deform and distort the field
lines. During this highly dynamical phase, viscous dissipation leads to
heating (preferentially of ions) of what would eventually become the solar
wind.

\section*{ACKNOWLEDGEMENTS}


Authors thank Abdus Salam International Centre for Theoretical Physics,
Trieste, Italy, where this work was started. The work of SMM was supported
by US DOE contract DE--FG03--96ER--54366. The work of KIN was supported by
grant 02--02--16199a from the Russia Fund of Fundamental Research (RFFR).
The work of NLS was supported by ISTC grant Project G-663.

\clearpage

\figcaption[fig1.eps]{ Magnetic field deformation caused by the
strong flows: (a) Initial distribution of the flow kinetic energy,
$|{\bf V_0}|_{max}(x=0)=600\,km/sec, \ n_0\sim 10^8\,cm^{-3},$ (b)
The evolution of the arcade--like magnetic structure  displayed for
3 distinct time--frames: \ $t = 0; \ 768 ; \ 1749\,sec$; the
structure had initially \ $B_{0max}(0,Z=0)=5\,G$. It is shown that
the strong shear is created in the central region of the structure
resulting in the plasmoid--type configuration; this configuration may
not lead to the particle escape. In this and later figures the
heights are measured from the Sun's surface.
\label{fig1}
}

\figcaption[fig2.eps]{  Magnetic field deformation caused by
interactions with the strong flow: (a)  Initial distribution of flow
kinetic energy, $|{\bf V_0}|_{max}(x=0)=600\,km/sec, \ n_0\sim
10^8\,cm^{-3}$; the structure had initially \ $B_{0max}(0,Z=0)=5\,G$,
(b) The evolution of the arcade--like magnetic structure  for 3
distinct time--frames: $t = 0; \ 768 ; \ 1749\,sec $. It is shown
that the strong shear is created in the central region of the
structure resulting in the well--type configuration; this
configuration too may not lead to the particle escape and even to the
cold prominence creation.
\label{fig2}
}

\figcaption[fig3.eps]{Dynamical Creation of a particle escape
channel
in a mgnetic structure of 2 identical arcades by a strong flow: (a)
For initial and boundary conditions : $B_{0max}=3\,G$, \ flow $|{\bf
V_0}|_{max}=920\,km/sec, \ n_0=2\cdot 10^7\,cm^{-3}$, background
plasma density $=5\cdot 10^6\,cm^{-3}$ at the height where the strong
flows can be found, (b) Plots for the vector potential $A$; the
density $n$, the temperature $T$,  and the speed $|{\bf V}|$ \  for 3
different time frames \ $t=973\,sec; \ 1988\,sec; \ 3048\,sec $. The
channel for particle escape may be clearly seen. Note: The shock seen
at the leading edge is an artifact due to the interaction of the flow
with the background plasma (necessary for the smooth working of the
simulation code).
\label{fig3}
}

\figcaption[fig4.eps]{ Evolution of (a) the density $n(x=0,z)$; (b)
the flow radial velocity $V_z(x=0,z)$; and (c) the temperature
$T(x=0,z)$ in the center of the escape channel of Fig.(3) along the
radial distance $z$. A sharp decrease in density and the accompanying
ballistic deceleration of the initial flow is revealed. It is also
seen that flow is concentrated practically along the axis of the
channel. $z$-- projection of the shock explained in Fig.3 maybe be
clearly seen.
\label{fig4}
}

\figcaption[fig5.eps]{ Inhomogeneous and divergent boundary,
temporary channel creation in a structure of four identical arcades
: (a) Boundary conditions for the 3 pulse initial flow (spatially non--uniform),
$|{\bf V_0}|_{max}=920\,km/sec \ , \ n_0=2\cdot 10^7\,cm^{-3}$, background
density $=5\cdot 10^6\,cm^{-3}$; (b) initial condition for $A$
(initial magnetic field \ $B_{0max}(x_0,Z=0)=4\,G$) ; \ (c)
dynamical evolution results at $t= 2335\,sec$ for the potential $A$,
the temperature $T$ and the speed $|{\bf V}|$ . It is seen that
deceleration is ballistic and the flow occupies practically the entire
region of the 4 initial arcades.
\label{fig5}
}

\end{document}